\documentclass[aps,prd,preprint,floatfix]{revtex4}
\usepackage{epsfig}
\usepackage{bm}
\usepackage{dcolumn}
\usepackage{latexsym}

\begin{document}
   
\preprint{\rightline{ANL-HEP-PR-07-81}}
   
\title{Lattice QCD at finite temperature and density in the phase-quenched
approximation}
   
\author{J.~B.~Kogut}\thanks{Supported in part by NSF grant NSF PHY03-04252.}
\affiliation{Department of Energy, Division of High Energy Physics, Washington,
DC 20585, USA}
 \author{\vspace{-0.2in}{\it and}}
\affiliation{Dept. of Physics -- TQHN, Univ. of Maryland, 82 Regents Dr.,
College Park, MD 20742, USA}
\author{D.~K.~Sinclair}\thanks{This work was supported in part by the U.S.
Department of Energy, Division of High Energy Physics, \\*[-0.1in]Contract 
DE-AC02-06CH11357.}
\affiliation{HEP Division, Argonne National Laboratory, 9700 South Cass Avenue,
Argonne, IL 60439, USA}
                                                                                
\begin{abstract}
QCD at a finite quark-number chemical potential $\mu$ has a complex fermion
determinant, which precludes its study by standard lattice QCD simulations.
We therefore simulate lattice QCD at finite $\mu$ in the phase-quenched
approximation, replacing the fermion determinant with its magnitude. (The
phase-quenched approximation can be considered as simulating at finite isospin
chemical potential $2\mu$ for $N_f/2$  $u$-type and $N_f/2$ $d$-type quark
flavours.) These simulations are used to study the finite temperature
transition for small $\mu$, where there is some evidence that the position
(and possibly the nature) of this transition are unchanged by this
approximation. We look for the expected critical endpoint for 3-flavour QCD.
Here, it had been argued that the critical point at zero $\mu$ would become
the critical endpoint at small $\mu$, for quark masses just above the critical
mass. Our simulations indicate that this does not happen, and there is no such
critical endpoint for small $\mu$. We discuss how we might adapt techniques
used for imaginary $\mu$ to improve the signal/noise ratio and strengthen our
conclusions, using results from relatively low statistics studies.

\end{abstract}

\maketitle

\section{Introduction}

Relativistic heavy-ion colliders allow one to study hadronic and nuclear matter
at high temperatures where it undergoes a transition to a quark-gluon plasma.
While the highest energy colliders (RHIC and the forthcoming heavy-ion program
at the LHC) study only the very low density regime where the baryon-number
density is too small to have much effect on the thermodynamics, lower energy
relativistic heavy-ion colliders can probe the region where baryon-number
density is appreciable.

For physical $u$, $d$ and $s$ quark masses, the finite temperature transition
at zero baryon-number density is predicted to be a rapid crossover rather than
a true phase transition \cite{Karsch:2001nf,deForcrand:2006pv,Aoki:2006we}. 
It is expected that, at high enough baryon-number 
densities, this transition will become first order. The point at which the
change from a crossover to a first-order transition occurs would be a critical
point, expected to be in the universality class of the 3-dimensional Ising
model. This critical point is referred to as a critical endpoint, and is
expected to be the most interesting feature of this intermediate density
regime of the QCD phase diagram.

While finite temperature QCD is straightforward (but tedious) to simulate
on the lattice, QCD at a finite quark-number chemical potential $\mu$ has
proved intractable. The reason is that at finite $\mu$ the fermion determinant
becomes complex, with a real part having an indefinite sign. Since all the
standard lattice QCD simulation methods rely on importance sampling, they fail
for such systems.

In the region of small $\mu$, close to the finite-temperature phase transition,
methods have been developed to circumvent this sign problem. These methods fall
into several classes. One such method involves simulating lattice QCD at a
carefully selected set of parameters where no such sign problem exists and
using the ratios of determinants to reweight to the region of interest
\cite{Fodor:2004nz}. Such
multiparameter reweighting only works provided there is significant overlap
between those configurations which are important for the chosen set of
parameters, and those which are important for the original set of parameters.
A second class of methods are those which rely on analyticity in $\mu$ or
related parameters. These include series expansion methods 
\cite{Allton:2002zi,Gavai:2003mf}, which expand the
Boltzmann weight and the observables as power series in $\mu$, calculating
the coefficients in simulations at zero $\mu$. Since the higher order
coefficients require the calculation of higher order fluctuation quantities,
this ultimately limits their utility. Other analyticity methods involve 
simulating in a domain of parameters such as at imaginary $\mu$, where there
is no sign problem, and analytically continuing the results to the desired
domain (in this case, real $\mu$), typically by fitting the results to a power 
series \cite{de Forcrand:2002ci,D'Elia:2004at}. There exist variants where
different parameters are used for the analytic continuation such as
\cite{Azcoiti:2004ri}.
Another way of avoiding the sign problem is to use canonical methods
\cite{Engels:1999tz,deForcrand:2006ec,Alexandru:2005ix}. 
Here the sign problem is encountered in Fourier transforming to obtain the 
canonical ensembles at fixed quark number.

We have adopted the alternative approach of ignoring the phase of the 
determinant and replacing the determinant by its magnitude. This can be thought
of as simulating QCD with $N_f/2$ $u$ type quarks and $N_f/2$ $d$ type quarks,
with a chemical potential $\mu_I=2\mu$ for isospin ($I_3$). ($N_f$ is the number
of quark flavours.) For low temperatures, there is a critical point
$\mu_I=\mu_c$ above which the system enters a superfluid phase, with a charged
pion condensate which breaks $I_3$ symmetry spontaneously
\cite{Son:2000xc,Hands:1999md,Kogut:2002zg}. At zero temperature $\mu_c=m_\pi$.
Since this phase does not exist for QCD at finite $\mu$, the phase-quenched
approximation breaks down at the boundary of this superfluid domain, if not
before. 

The Taylor series calculations of the Bielefeld-Swansea collaboration
\cite{Allton:2002zi,Ejiri:2004yw} showed evidence that the $\mu$ dependence of
the transition temperature $T_c$ for full 2-flavour QCD was similar if not
identical to that of the phase-quenched approximation (finite isospin chemical
potential), at small $\mu$. The $\mu$ dependence of this transition for full
2-flavour QCD, obtained from the imaginary quark-number chemical potential
simulations of de Forcrand and Philipsen, \cite{de Forcrand:2002ci} was
consistent with being identical that observed in our direct simulation of the
phase-quenched theory \cite{Kogut:2004zg}. A random matrix model of 2-flavour
QCD at finite temperature and chemical potentials also predicts that the
dependence of $T_c$ on quark-number and isospin chemical potentials should be
identical for $\mu < m_\pi/2$ \cite{Klein:2003fy}. In addition
Nambu-Jona-Lasinio models for QCD have transition temperatures which exhibit
the same dependence on quark-number and isospin chemical potentials for $\mu <
m_\pi/2$ \cite{Toublan:2003tt,Barducci:2005ut}. This strongly suggests that
the $\mu$ dependence of $T_c$ is the same for phase-quenched and full QCD for
small $\mu$. We shall indicate later that the simulations discussed in this
paper are consistent with this assumption. There is, however, one lattice
result which contradicts this assumption. The 3-flavour calculations of the
Bielefeld-Swansea collaboration indicate that while the $T_c$ dependence on
quark-number and isospin chemical potentials are consistent at larger quark
masses, they are not at small quark masses
\cite{Karsch:2003va,Schmidt:2004ke}. However, because of large statistical
errors, the observed difference in slopes was less than two standard
deviations. In addition, these simulations were performed using the R
algorithm, which could potentially introduce larger than expected
updating errors, due to the discretization of molecular-dynamics `time', in the 
fluctuation quantities used to obtain these results.


For 3-flavour QCD at zero chemical potentials, the finite temperature 
transition is first order at small quark mass $m$. For larger $m$ the 
transition softens to a crossover with no phase transition. At $m=m_c$, where
the nature of the transition changes, the finite-temperature transition is
a critical point in the universality class of the 3-dimensional Ising model
\cite{Karsch:2001nf}.
Similar behaviour is seen for $2+1$-flavour QCD, and for the physical $u$, $d$ 
and $s$ quark masses, the transition is predicted to be a crossover
\cite{deForcrand:2006pv,Aoki:2006we}. It has
been suggested that $m_c$ would increase with increasing $\mu$, becoming the
critical endpoint. If so it should be possible to tune this endpoint to be
as close to $\mu=0$ as desired by choosing $m$ just above $m_c$.

Hence we simulate 3-flavour lattice QCD at $\mu_I < m_\pi$, for several masses
close to $m_c$, and determine the nature of the finite-temperature phase 
transition using fourth-order Binder cumulants for the chiral condensate. 
For these studies we use simulations on $8^3 \times 4$, $12^3 \times 4$ and
$16^3 \times 4$ lattices. Our simulations indicate that there is no critical
endpoint for $m > m_c(0)$, and $m_c(\mu_I)$ actually decreases (slowly) with
increasing $\mu$. Preliminary results from these simulations have been
reported at various conferences, the most recent being Lattice2007 
\cite{Sinclair:2007ce}. This absence of the expected critical endpoint
at small $\mu$($\mu_I$) has been observed by de Forcrand and Philipsen using
analytic continuation from imaginary $\mu$ \cite{deForcrand:2006pv}. 
Our simulations use the exact RHMC algorithm \cite{Clark:2006wp}, 
since in the inexact hybrid molecular-dynamics used in earlier
simulations, the Binder cumulant had such strong $dt^2$ dependence as to lead
to incorrect conclusions as to the nature of the transition \cite{Kogut:2006jg}.

The relatively weak dependence of the Binder cumulant on $\mu_I^2$ and the
statistical errors in determining it, even in high statistics runs, make it
difficult to determine the sign of the slope $d B_4/d \mu_I^2$ and hence 
$d m_c/d \mu_I^2$ with certainty. Similar difficulties occurred with the
methods of de Forcrand and Philipsen, but they were able to calculate the
slope directly with much higher precision, using reweighting methods 
\cite{deForcrand:2007rq}. We have performed studies which indicate that
similar methods should work for the phase-quenched simulations. However, on
the larger lattices we use, it is unclear whether these methods will be
significantly more efficient than simply increasing statistics. As yet, we
have insufficient statistics to achieve results for the slope of the Binder
cumulant. However, we are already able to determine the slope of $\beta_c$. 

Section~2 describes phase-quenched lattice QCD. In section~3 we describe our
simulations and results. Exploratory studies of reweighting techniques are
described in section~4. Section~5 is devoted to discussions and conclusions.

\section{Phase-quenched lattice QCD}

Phase-quenched lattice QCD with eight staggered quark flavours (or two 
staggered quark fields, each with four `tastes') has the fermion action
\begin{equation}
S_f=\sum_{sites} \left[\bar{\chi}[D\!\!\!\!/(\frac{1}{2}\tau_3\mu_I)+m \right]
                                                                          \chi
\end{equation}
where $D\!\!\!\!/(\frac{1}{2}\tau_3\mu_I)$ is the standard staggered quark
transcription of $D\!\!\!\!/$ with the links in the $+t$ direction multiplied
by $\exp(\frac{1}{2}\tau_3\mu_I)$ and those in the $-t$ direction multiplied by
$\exp(-\frac{1}{2}\tau_3\mu_I)$. Since we are performing simulations outside
of the superfluid phase, the explicit symmetry-breaking interaction of our
earlier studies is unnecessary.

To simulate $N_f$ flavours using the RHMC algorithm, this is replaced by the
pseudo-fermion action
\begin{equation}
S_{pf}=p_\psi^\dag {\cal M}^{-N_f/8} p_\psi
\end{equation}
where $p_\psi$ are the momenta conjugate to the pseudo-fermion field $\psi$%
\footnote{We choose to call these momenta rather than fields, since we leave
open the possibility of adding a function of $\psi$ to the action, which does
not change the physics, but destroys the partial integrability of the
equations-of-motion along a trajectory. This just means adding a familiar
potential term, whereas adding terms of higher order in the momenta is
somewhat less familiar.}. Here,
\begin{equation}
{\cal M} = [D\!\!\!\!/(\frac{1}{2}\mu_I)+m]^\dag 
           [D\!\!\!\!/(\frac{1}{2}\mu_I)+m].
\end{equation}
In the RHMC algorithm ${\cal M}^{-N_f/8}$ (and ${\cal M}^{\pm N_f/16}$) are
replaced by rational approximations, using a speculative lower bound
\cite{Kogut:2006jg}. It is interesting to note that these rational
approximations provide similar infrared protection to what a symmetry-breaking
interaction would give.

For 8 flavours, and indeed for any even number of flavours, $\mu_I$ has the
interpretation of an isospin chemical potential, for a theory with $N_f/2$
$u$-type quarks and $N_f/2$ $d$-type quarks. Since we are interested in this 
phase-quenched theory as an approximation to QCD with a quark-number chemical
potential $\mu=\mu_I/2$, we are free to choose any integral $N_f$. In fact we
shall work with $N_f=3$. 

As we have shown in earlier work, the Binder cumulant which is used to extract
the nature of the finite-temperature transition is very sensitive to the
updating increment $dt$ in the older, inexact, hybrid molecular-dynamics (R)
algorithm \cite{Kogut:2006jg}. This is the principal reason that we have
switched to the RHMC algorithm.

As mentioned in the introduction, such theories are known to undergo a phase
transition to a superfluid phase with a charged pion condensate and
orthogonal charged pion excitations which are true Goldstone bosons at low
temperatures, as $\mu_I$ is increased. At zero temperature this transition 
occurs at $\mu_I=\mu_c=m_\pi$. At high enough temperatures the system should
be in the quark-gluon phase for all $\mu_I$, and no such transition is expected.
 
\section{Simulations and results}

We perform simulations of 3-flavour lattice QCD at finite $\mu_I$ and 
temperature on $8^3 \times 4$, $12^3 \times 4$, and $16^3 \times 4$ lattices.
We use rational approximations to ${\cal M}^{-3/8}$ and ${\cal M}^{\pm 3/16}$
in these RHMC simulations which are valid provided the spectrum of ${\cal M}$
is in the range $[1 \times 10^{-4}, 25]$. (2 runs were made using smaller
speculative lower bounds for testing purposes.) We performed runs with quark
masses $m=0.02$, $m=0.025$, $m=0.03$ and $m=0.035$ on $8^3 \times 4$ and $12^3
\times 4$ lattices. At the lowest mass, we only ran simulations for $\mu_I=0$.
For the other 3 masses we ran simulations at $\mu_I=0$, $\mu_I=0.2$ and
$\mu_I=0.3$. In addition, we ran simulations on $16^3 \times 4$ lattices at
$m=0.03$ at all 3 $\mu_I$~s and at $m=0.025$ with $\mu_I=0$. The masses are
chosen such that the lower 2 masses lie below $m_c$ and the higher 2 masses
lie above $m_c$. The choice of $\mu_I$ values is to cover the region 
$0 \le \mu_I < m_\pi$, where $m_\pi$ is estimated to lie in the range 
$0.4 \lesssim m_\pi < 0.5$ for $0.025 \le m \le 0.035$. (This comes from
measurements of $m_\pi$ for $m=0.03$, $\mu_I=0$ at $\beta=5.10$ and 
$\beta=5.15$, which bracket the region of interest, on $8^3 \times 16$ and 
$12^3 \times 24$ lattices. Estimates for other $m$ values were made using PCAC.)

For our $12^3 \times 4$ simulations, where we have the highest statistics, we
have run for 300,000 length-1 trajectories for each of 4 (or more) $\beta$
values close enough to the transition to access this transition using
Ferrenberg-Swendsen reweighting 
in $\beta$, at each $(m,\mu_I)$. For the
$8^3 \times 4$ and $16^3 \times 4$ simulations we have performed 300,000
trajectory runs at each of 2 $\beta$s at each $(m,\mu_I)$. We have made 5 
independent stochastic estimates of the chiral condensate $\bar{\psi}\psi$ and 
the isospin density $j_0^3 = \partial S_f/\partial \mu_I$ after each 
trajectory, to enable us to make unbiased estimates of the susceptibilities
and Binder cumulants. 

For any observable $X$, the susceptibility $\chi_X$ is defined by
\begin{equation}
\chi_X = \frac{V}{T} \langle \overline{X}^2 - \langle \overline{X} \rangle^2
                                                                   \rangle,
\end{equation}
where $V$ is the spatial volume and $T=1/N_t$ is the temperature. The 
overlining of $X$ indicates that these are lattice averaged quantities. The
fourth-order Binder cumulant for $X$ is defined by 
\begin{equation}
B_4 = {\langle (\overline{X} - \langle \overline{X} \rangle)^4 \rangle \over
       \langle (\overline{X} - \langle \overline{X} \rangle)^2 \rangle^2}
\end{equation} 
\cite{binder}.
These quantities are measured at the value $\beta=\beta_0$
of the simulation and extrapolated to nearby $\beta$s, by Ferrenberg-Swendsen
reweighting:
\begin{equation}
\langle X \rangle_\beta = {\langle \exp[-6(V/T)(\beta-\beta_0)S_\Box] X \rangle
                                                                     _{\beta_0}
                    \over  \langle \exp[-6(V/T)(\beta-\beta_0)S_\Box] \rangle
                                                                     _{\beta_0}}
\end{equation} 
\cite{Ferrenberg:yz} where 
\begin{equation}
S_\Box = 1-\frac{1}{3}{\rm Re\,Tr}_\Box UUUU
\end{equation}
appropriately averaged over the lattice and over plaquette orientations. The
position of the transition, $\beta_c$, can be estimated as that of the peak of
the susceptibilities, or the minimum of the Binder cumulants. We have noticed
that the measured $\beta_c$s from the susceptibilities of various observables
and from the Binder cumulants are close.

The Binder cumulant for the chiral condensate is used to probe the nature
(as well as the position) of the transition. In the infinite volume limit,
$B_4=3$ at a crossover, $B_4=1$ at a first-order transition and $B_4=1.604(1)$
at a 3-dimensional Ising critical point. If there indeed were a critical
endpoint, for $m > m_c(0)$ $B_4$ would start at a value above the Ising value
for $\mu_I=0$ (close to $3$ for really large lattices) and decrease, passing
through a value close to the Ising value at the critical endpoint, eventually
approaching $1$ for large $\mu_I$. On large enough lattices, finite-size
scaling predicts that lines of $B_4$ versus $\mu_I$ for different lattice
sizes will cross at the Ising value. Similarly lines of $B_4$ versus $m$
for different size lattices will cross at the Ising value as $m$ is varied. 

\begin{figure}[htb]
\epsfxsize=3.0in
\epsffile{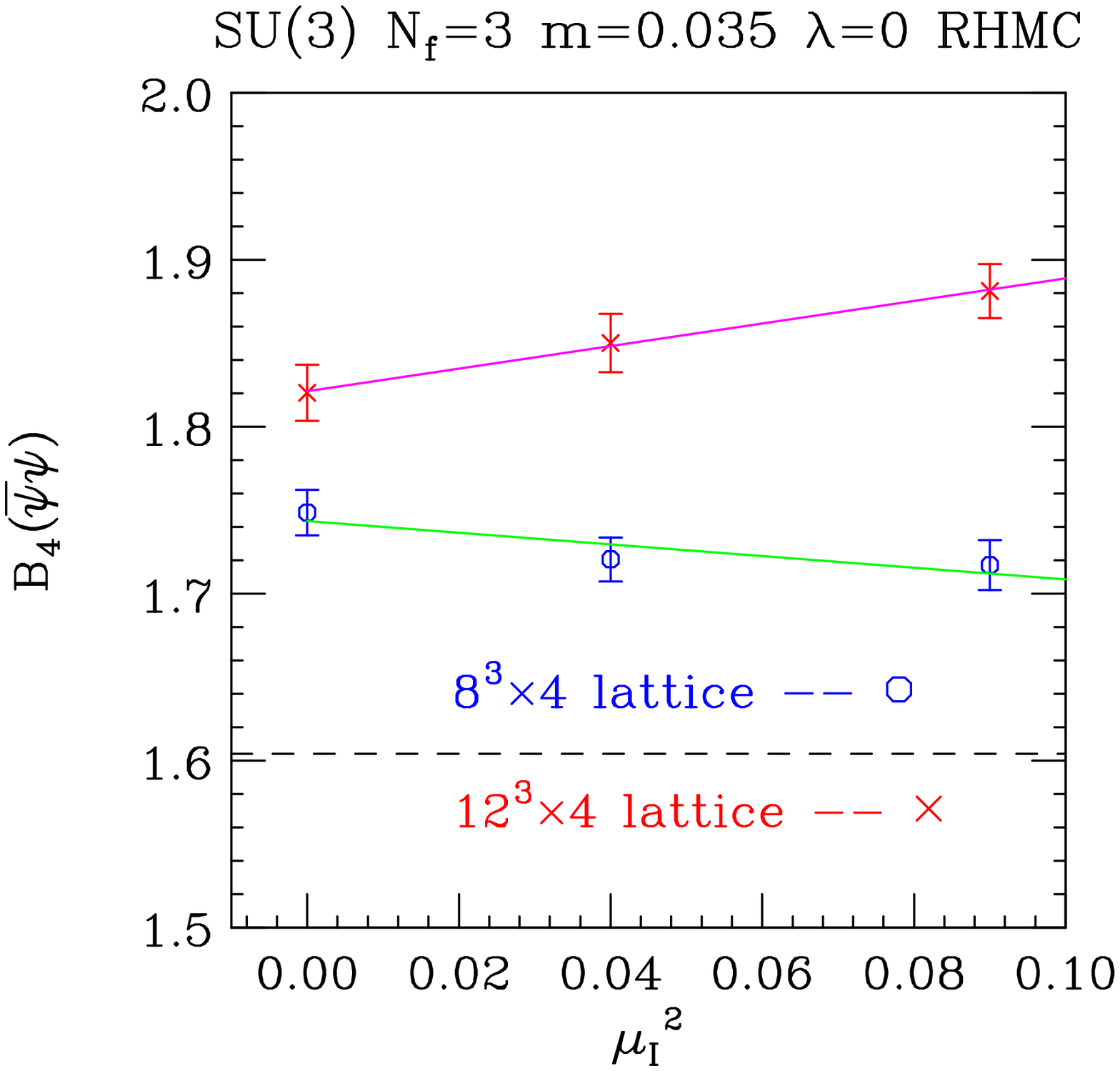}
\epsfxsize=3.0in
\epsffile{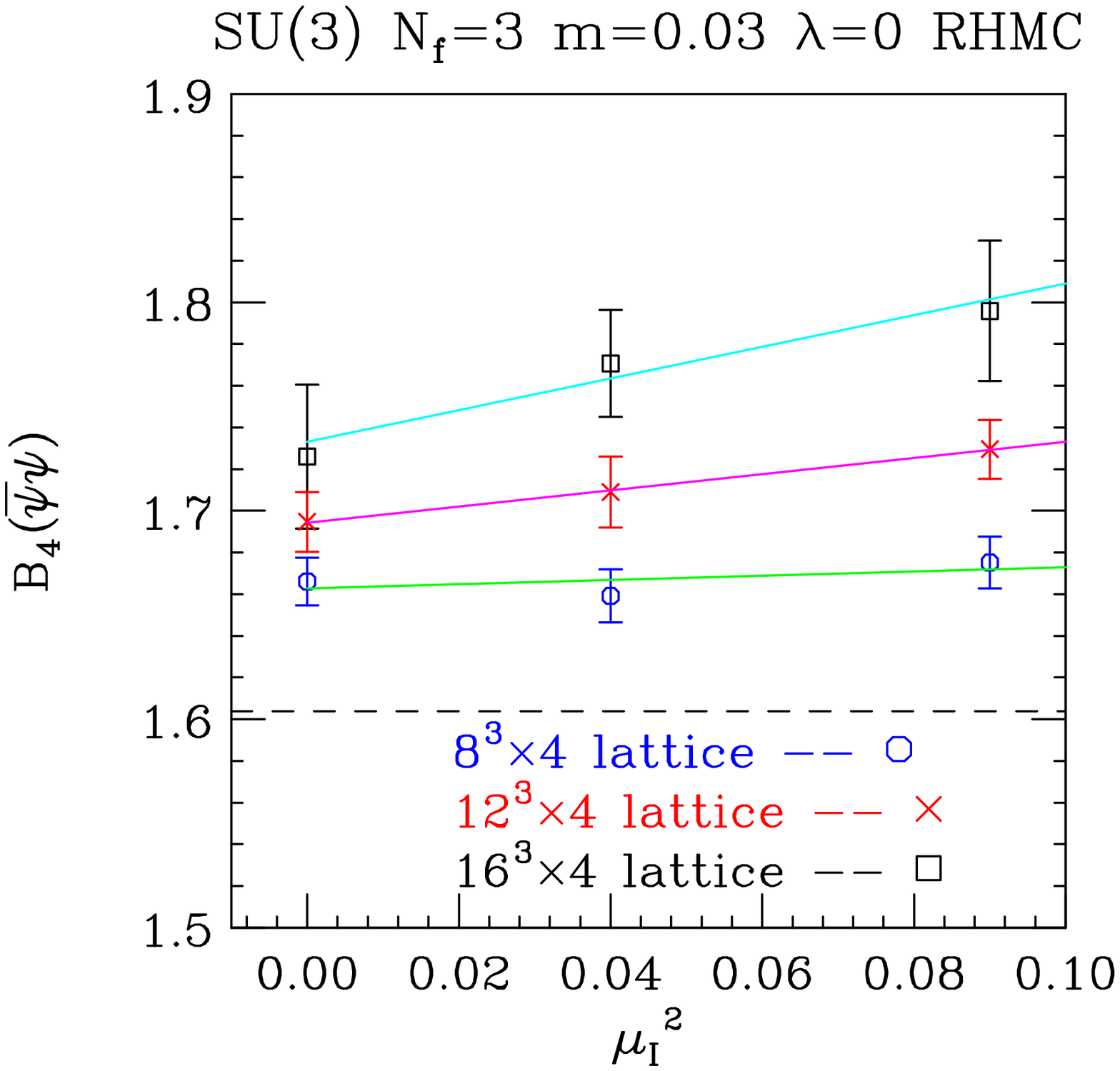}
\vspace{0.25in}
\epsfxsize=3.0in
\centerline{\epsffile{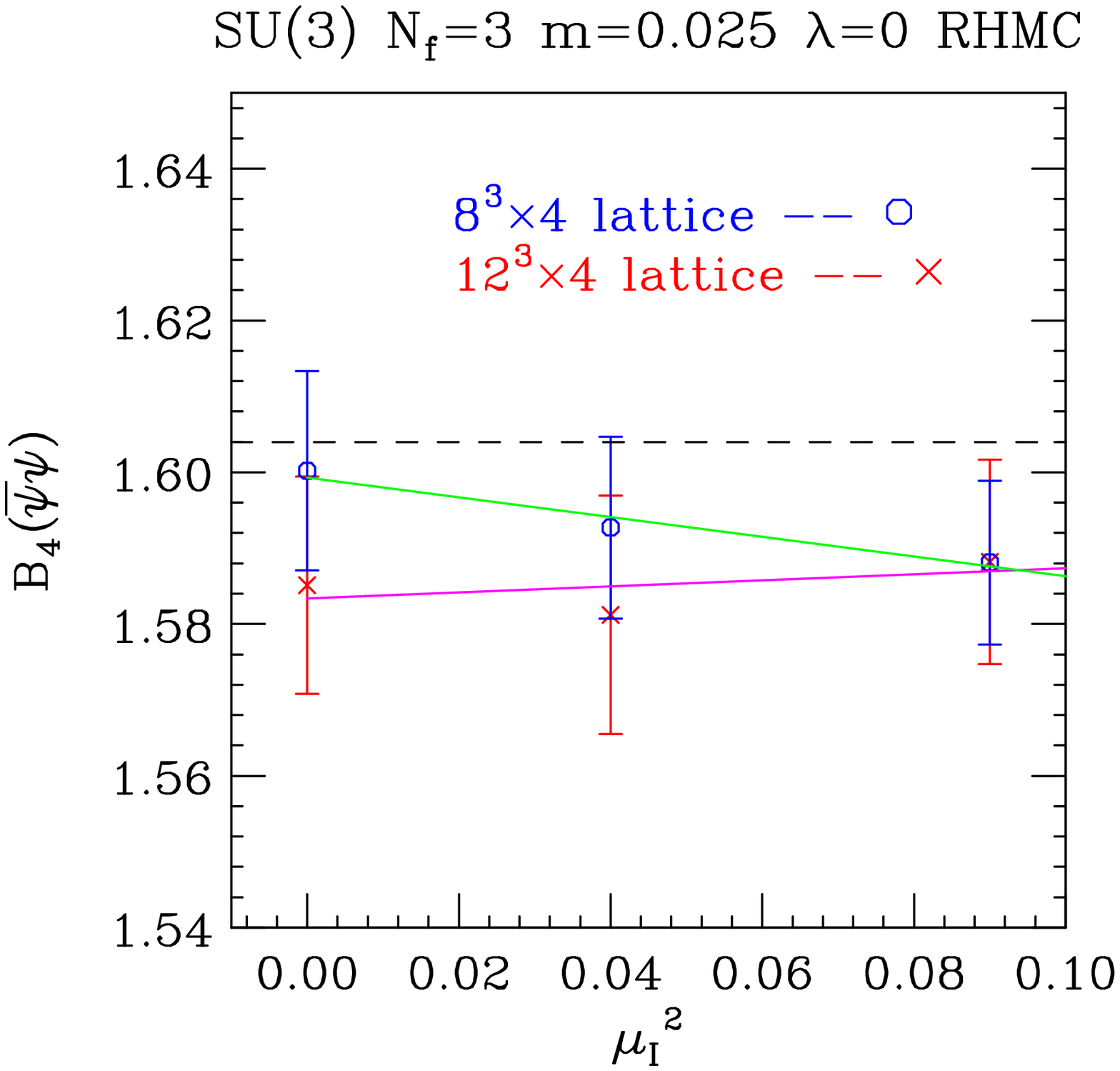}}
\caption{Graphs showing the $\mu_I^2$ dependence of the Binder cumulants for
the chiral condensate $\bar{\psi}\psi$ at the transition:
a) for $m=0.035$, b) for $m=0.03$, c) for $m=0.025$. The dashed line is at the
Ising value.}
\label{fig:B4vsmu}
\end{figure}

In figure~\ref{fig:B4vsmu} we plot the Binder cumulants at the transition point
as functions of $\mu_I^2$ for $m=0.025$, $m=0.03$ and $m=0.035$ for the
various lattice sizes. For $m=0.035$ $B_4$ on the $12^3 \times 4$ lattice
starts at a value significantly above that for an Ising critical point and
appears to increase with increasing $\mu_I$, and hence shows no evidence for
a critical endpoint the slope of this straight line fit is $0.68(26)$. 
Similarly, for $m=0.03$, $B4$ starts above the Ising value and increases with
increasing $\mu_I^2$ on the $12^3 \times 4$ and $16^3 \times 4$ lattices. For
the $12^3 \times 4$ lattice the slope is $0.39(22)$, and for the $16^3 \times
4$ lattice, this slope is $0.76(53)$. For $m=0.025$, there is no evidence for
any $\mu_I^2$ dependence for $B_4$ on the $12^3 \times 4$ lattice and it 
remains below the Ising value for the range of $\mu_I^2$ considered. We note
that the $8^3 \times 4$ slopes appear negative for $m=0.035$ and $m=0.025$ and
positive for $m=0.03$, which we interpret as meaning that we have insufficient
statistics to determine the very small slopes of the $8^3 \times 4$ lines.

None of the slopes we have measured is much more than $2\frac{1}{2}$ standard
deviations from zero. However, the fact that the two $12^3 \times 4$ slopes and
the one $16^3 \times 4$ slope for $m > m_c$ are all positive makes it less
likely that this is a statistical fluctuation. For $m < m_c$ we can draw no
conclusions.

\begin{figure}[htb]
\epsfxsize=3.0in
\epsffile{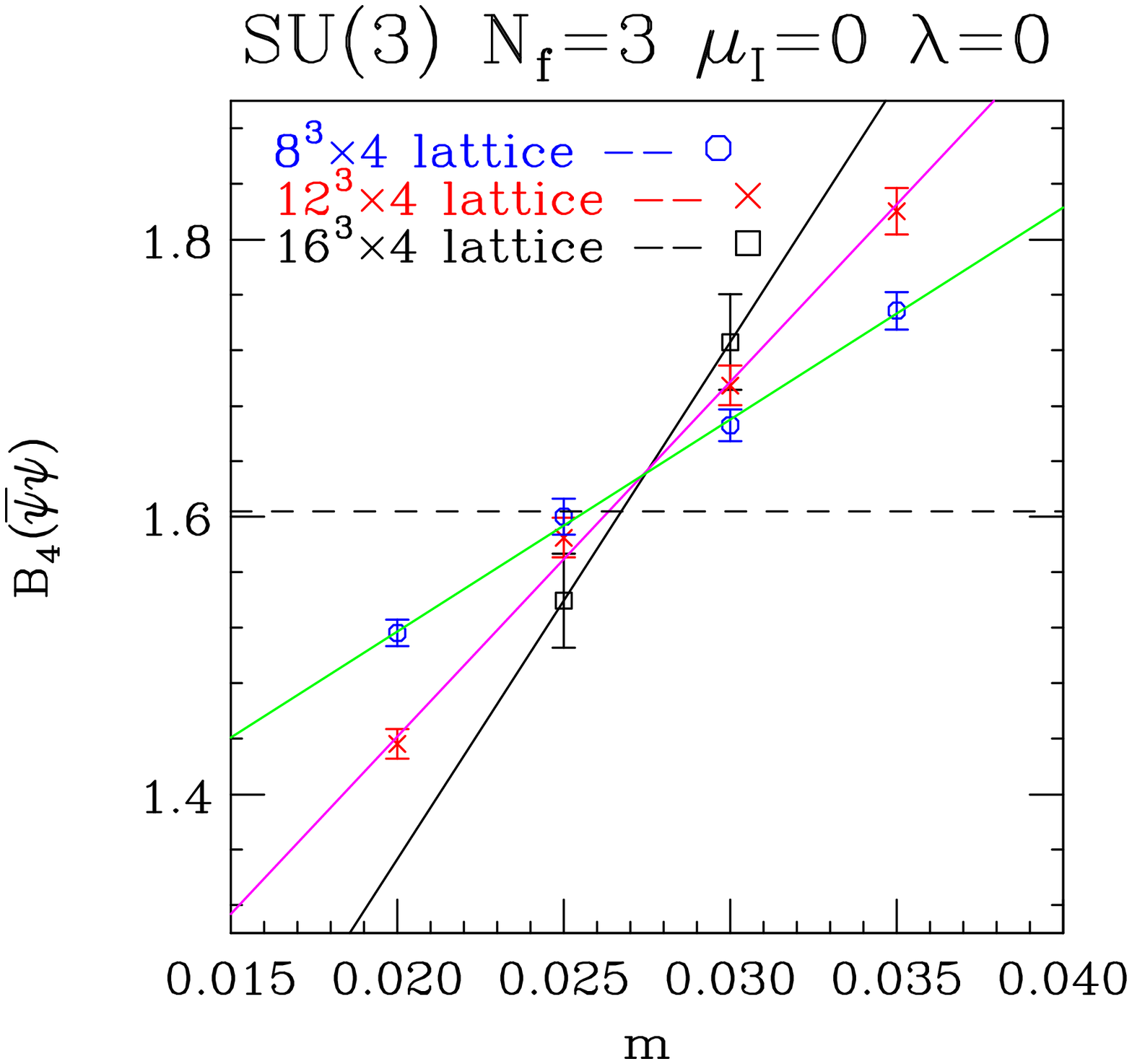}
\epsfxsize=3.0in
\epsffile{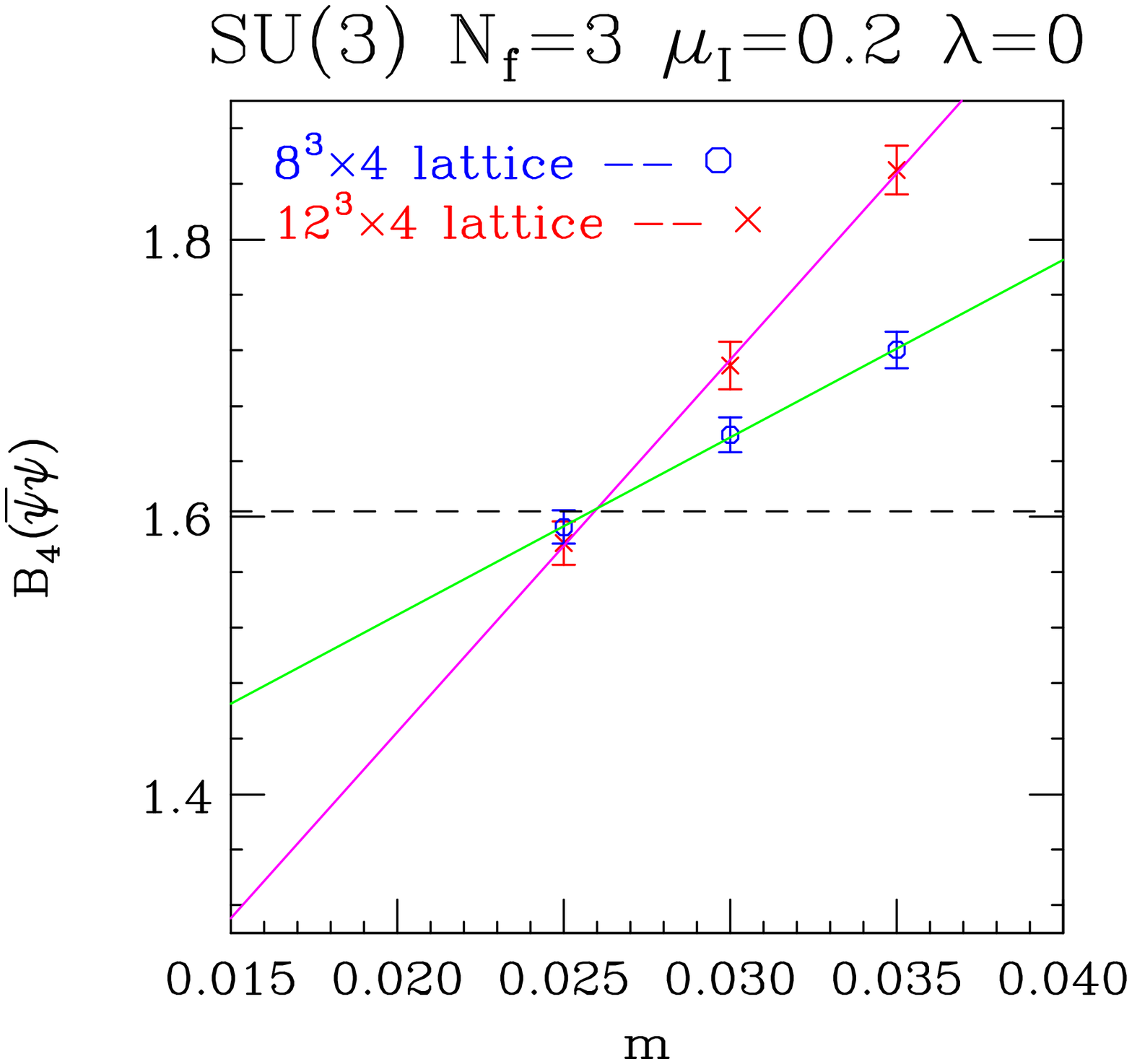}
\vspace{0.25in}
\epsfxsize=3.0in                                  
\centerline{\epsffile{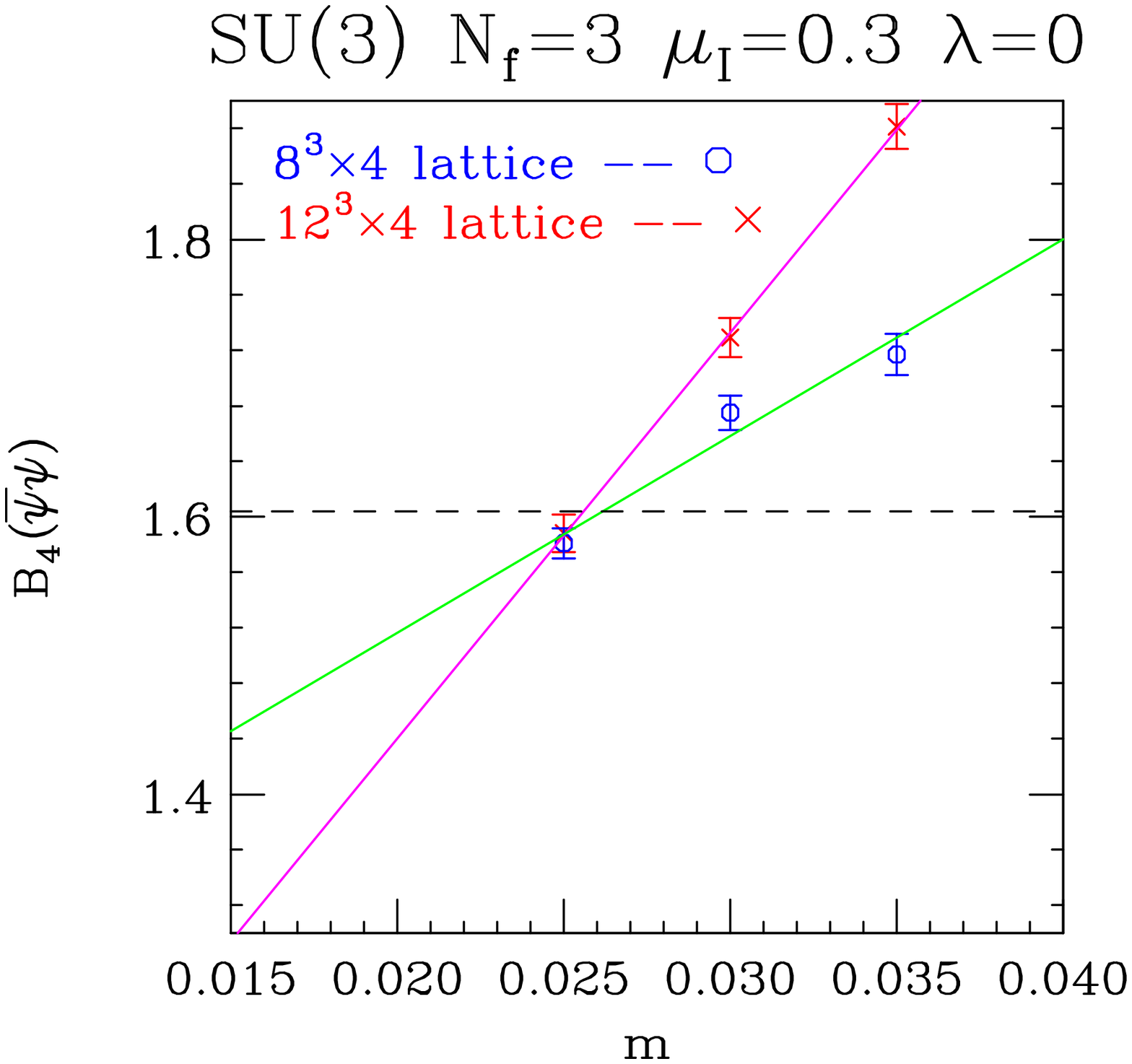}}
\caption{Graphs showing the $m$ dependence of the Binder cumulants for
the chiral condensate $\bar{\psi}\psi$ at the transition:                   
a) for $\mu_I=0$, b) for $\mu_I=0.2$, c) for $\mu_I=0.3$. The dashed line is
at the Ising value.}
\label{fig:B4vsm}                                       
\end{figure}

We now turn our attention to the mass dependence of $B_4$ at fixed $\mu_I$
values. Figure~\ref{fig:B4vsm} shows the $m$ dependence of $B_4$ for $\mu_I=0$,
$\mu_I=0.2$ and $\mu_I=0.3$.
First we note that the intersection of the curves for the different lattice
sizes intersect at $B_4$ close to its value for the 3-dimensional Ising model.
This is strong evidence that this critical point {\it is} in the universality
class of the 3-dimensional Ising model, as predicted. We therefore use the
masses for which the $12^3 \times 4$ Binder cumulants achieve the Ising value
as our estimate for the position of the critical point for the $\mu_I$ under
consideration. We get $m_c(0)=0.0265(3)$, $m_c(0.2)=0.0259(5)$ and 
$m_c(0.3)=0.0256(4)$. A straight line fit yields
\begin{equation}
m_c(\mu_I) = 0.0265(3) - 0.10(6)\,\mu_I^2 .
\end{equation}
This suggests that $m_c$ decreases with increasing $\mu_I$, rather than 
increasing as would be needed for a critical endpoint. Note also that if we
were to use the intersections of the curves for different lattice sizes as
our estimates for $m_c$, this would slightly increase our estimate for $m_c(0)$,
slightly decrease our estimate of $m_c(0.3)$ and leave our estimate of 
$m_c(0.2)$ essentially unchanged. This would make the slope even more negative.
In addition, since the transition temperature decreases with increasing $\mu_I$,
$m_c$ in physical units will decrease slightly faster than the $m_c$ in lattice
units, which we have presented here.

We have also examined the Binder cumulants for the isospin density $j_0^3$,
and find that these are consistent with those for the corresponding chiral
condensates. However, since our estimates for $j_0^3$ are much noisier, the
errors on $B_4(j_0^3)$ are considerably larger than those for 
$B_4(\bar{\psi}\psi)$, which makes them less useful. The Binder cumulants
for the plaquettes are appreciably larger, which is expected, since these
should be a reasonable approximation to the energy-like order parameter whose
Binder cumulant would approach $3$, even in the first-order regime and at the
critical point. 

The critical behaviour of this theory will be described by an effective 
Hamiltonian which is a linear combination of 3 fields, each of which has 
finite size scaling properties with one critical exponent. One such field
has the scaling behaviour of a `magnetization', a second that of an `energy'
and the third that of a `density'. Each will be given as a linear combination
of $\bar{\psi}\psi$, $S_g=(1-\frac{1}{3}{\rm Tr}_\Box UUUU)$ and $j_0^3$.
The simpler case at $\mu_I=0$, where there are only 2 fields to consider, has 
been studied in reference \cite{Karsch:2001nf}. In that paper they were able
to find simple expressions for the two `mixing' parameters. In our case there
are six such `mixing' parameters, we have been unable to find the six equations
required to determine these coefficients. If we were able to obtain these
eigenmodes of the renormalization group, the Binder cumulant of the 
`magnetization' order parameter would pass through the Ising value, once our
lattice is large enough that subdominant terms in finite size scaling
relation could be ignored. Until we can find such relationships we use the
fact that, on large enough lattices, the magnetization component of the chiral
condensate will dominate and its Binder cumulant will approach that of this
eigenmode. The fact that the Binder cumulants for the chiral condensate cross
close to the Ising value is evidence that this field is not strongly affected
by contamination from non-magnetic eigenmodes, on the lattice sizes we use. 

\begin{figure}[htb]
\epsfxsize=3.0in
\epsffile{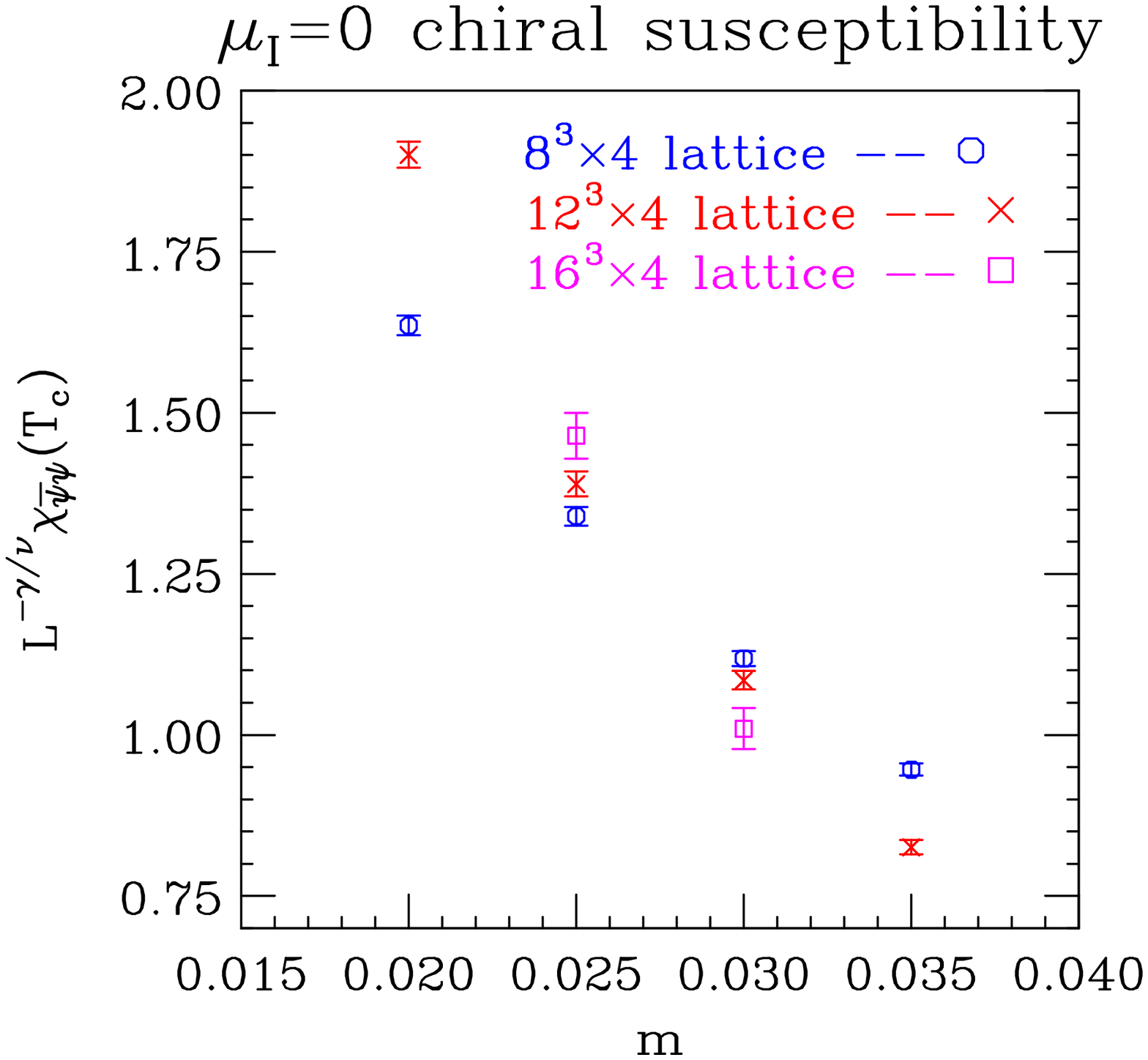}
\epsfxsize=3.0in
\epsffile{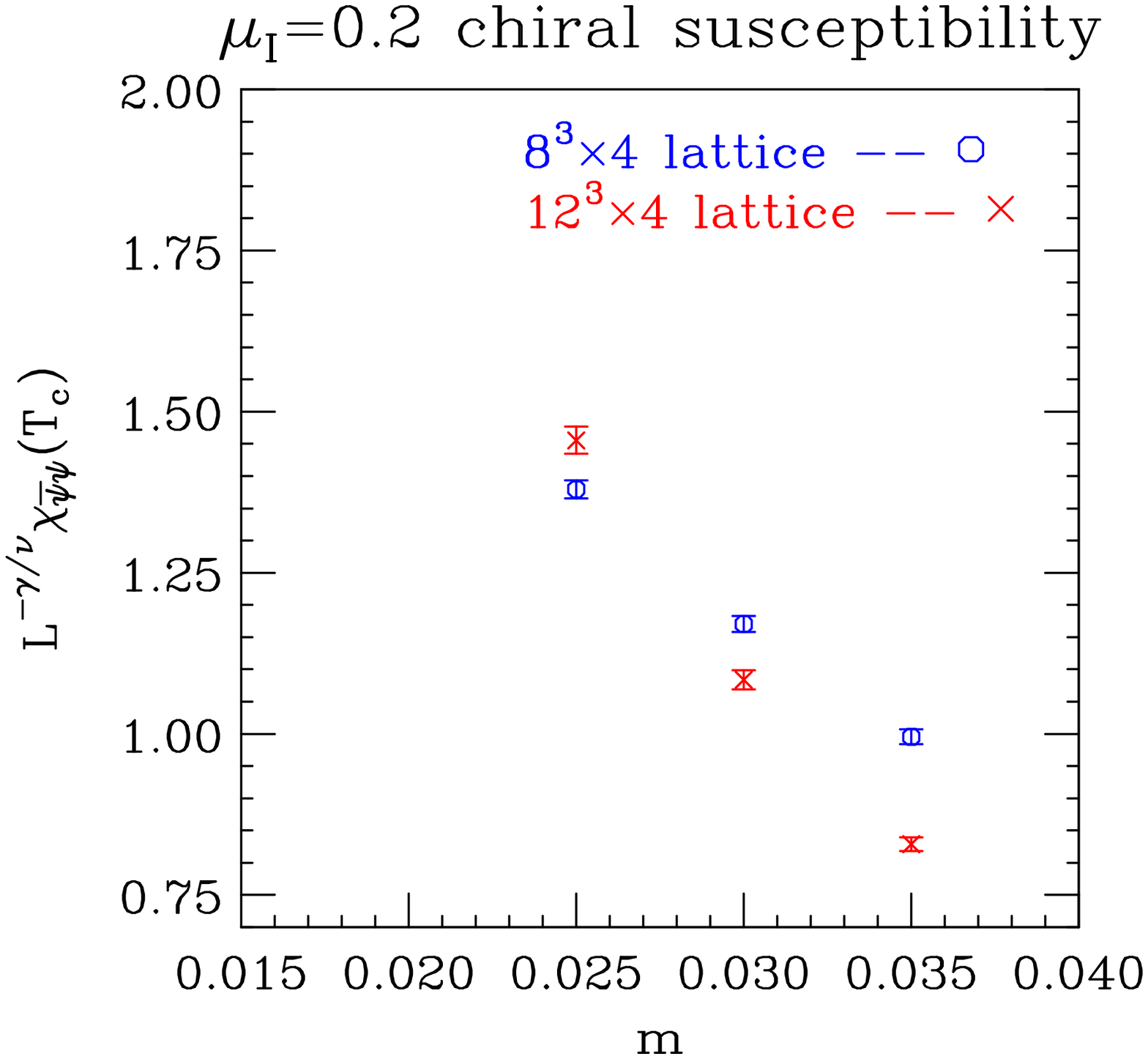}
\vspace{0.25in}
\epsfxsize=3.0in
\centerline{\epsffile{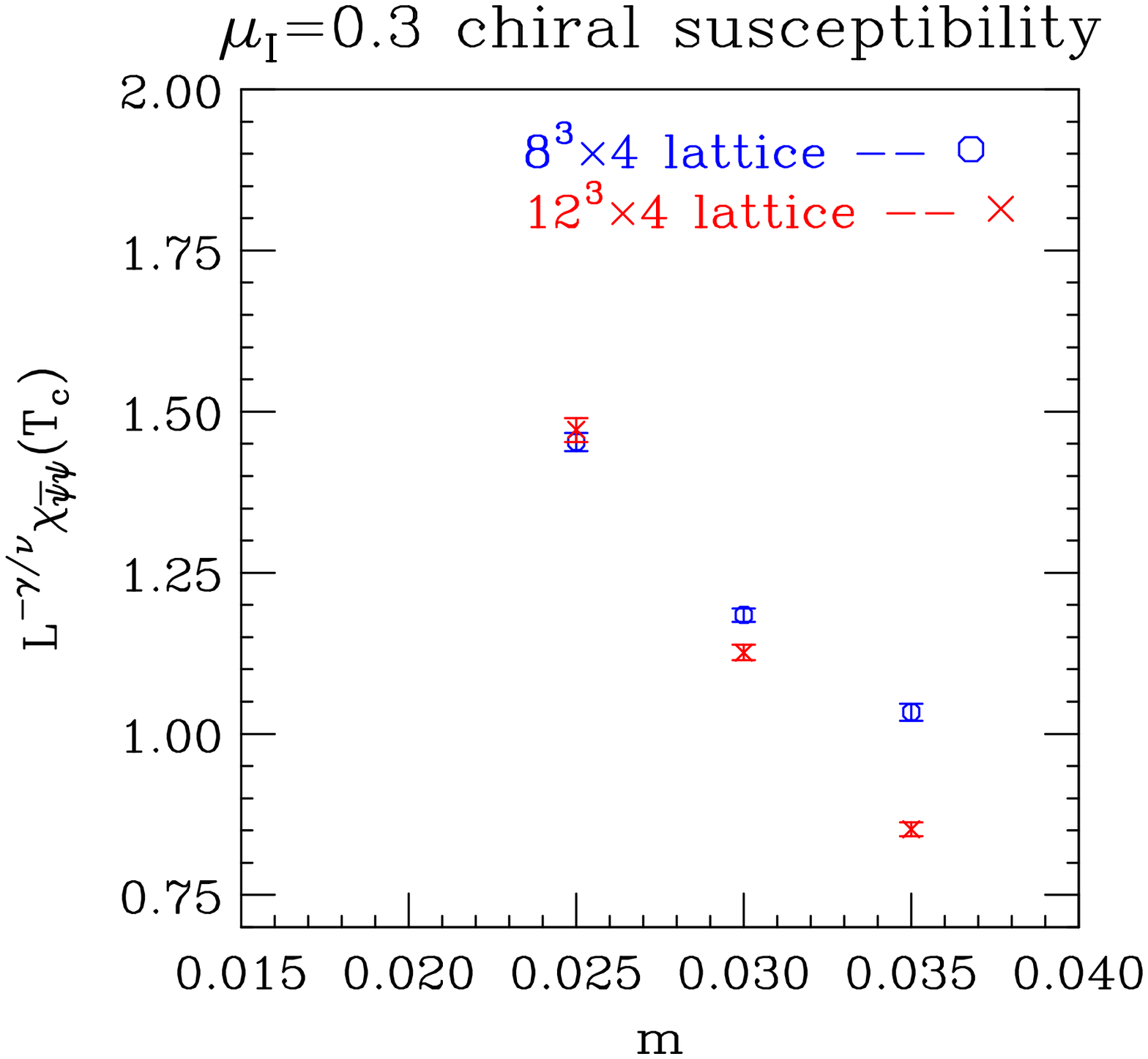}}
\caption{The rescaled chiral susceptibilities $\chi_{\bar{\psi}\psi}$ as 
functions of $m$: a) for $\mu_I=0$, b) for $\mu_I=0.2$, c) for $\mu_I=0.3$.}
\label{fig:fss}
\end{figure}

Using Ferrenberg-Swendsen reweighting again, we calculate the chiral 
susceptibilities and measure the positions and values of the peaks. We observe
that the positions of these peaks are very close to the minima of the Binder
cumulants. Finite size scaling tells us that, at the critical point,
\begin{equation}
\chi_{\bar{\psi}\psi}(L,T_c) = L^\frac{\gamma}{\nu} \tilde{\chi}
\end{equation}
where $L$ is the spatial extent of the lattice and $T_c$ is the critical
temperature. Hence if we plot 
$L^{-\frac{\gamma}{\nu}}\chi_{\bar{\psi}\psi}(L,T_c)$
as functions of $m$ for different $L$ values, the curves should cross at the
critical point. In figure~\ref{fig:fss} we plot this quantity for $L=8$ and
$L=12$, for each of our $\mu_I$ values. Here we have taken $\gamma=1.237$ and
$\nu=0.630$ as the required critical indices for the 3-dimensional Ising model.

Because it is clear that the points on this graph do not fall on straight
lines and the curves for different lattice sizes cross at rather shallow
angles, a quantitative estimate for the position of the crossing would be
difficult to obtain. What is clear is that the curves for the different
lattice sizes cross somewhere between $m=0.25$ and $m=0.3$ for $\mu_I=0$ and
$\mu_I=0.2$ and close to $m=0.25$ for $\mu_I=0.3$, which is consistent
with our estimates of $m_c(\mu_I)$ from Binder cumulants.

As well as trying to determine the nature of the finite temperature transition
as a function of $\mu_I$, and measuring observables and susceptibilities, the
positions of the minima in the Binder cumulants, and the positions of the
maxima in the various susceptibilities yield predictions for $\beta_c$ the
transition $\beta$ values. The $\mu_I$ dependence of $\beta_c$ will ultimately
yield the $\mu_I$ dependence of the transition temperature $T_c$. This not only
requires that we know $T_c$ at $\mu_I=0$, which we can obtain from the 
numerous measurements by other groups, but it also requires that we know the
renormalization group running of $\beta$ with lattice spacing $a$. On the
coarse lattices we use, 2-loop perturbative running of the coupling constant
which has been used earlier, is clearly suspect. Hence we present only the
$\mu_I$ dependence of $\beta_c$ in this paper. Associated with our present
simulations aimed at determining the equation-of-state for phased-quenched
(lattice) QCD, we will measure the running of $\beta$ directly with the 
same action and masses as are used here, on zero temperature lattices. At that 
time we will be able to predict the $\mu_I$ dependence of $T_c$.

\begin{figure}[htb]
\epsfxsize=6in
\centerline{\epsffile{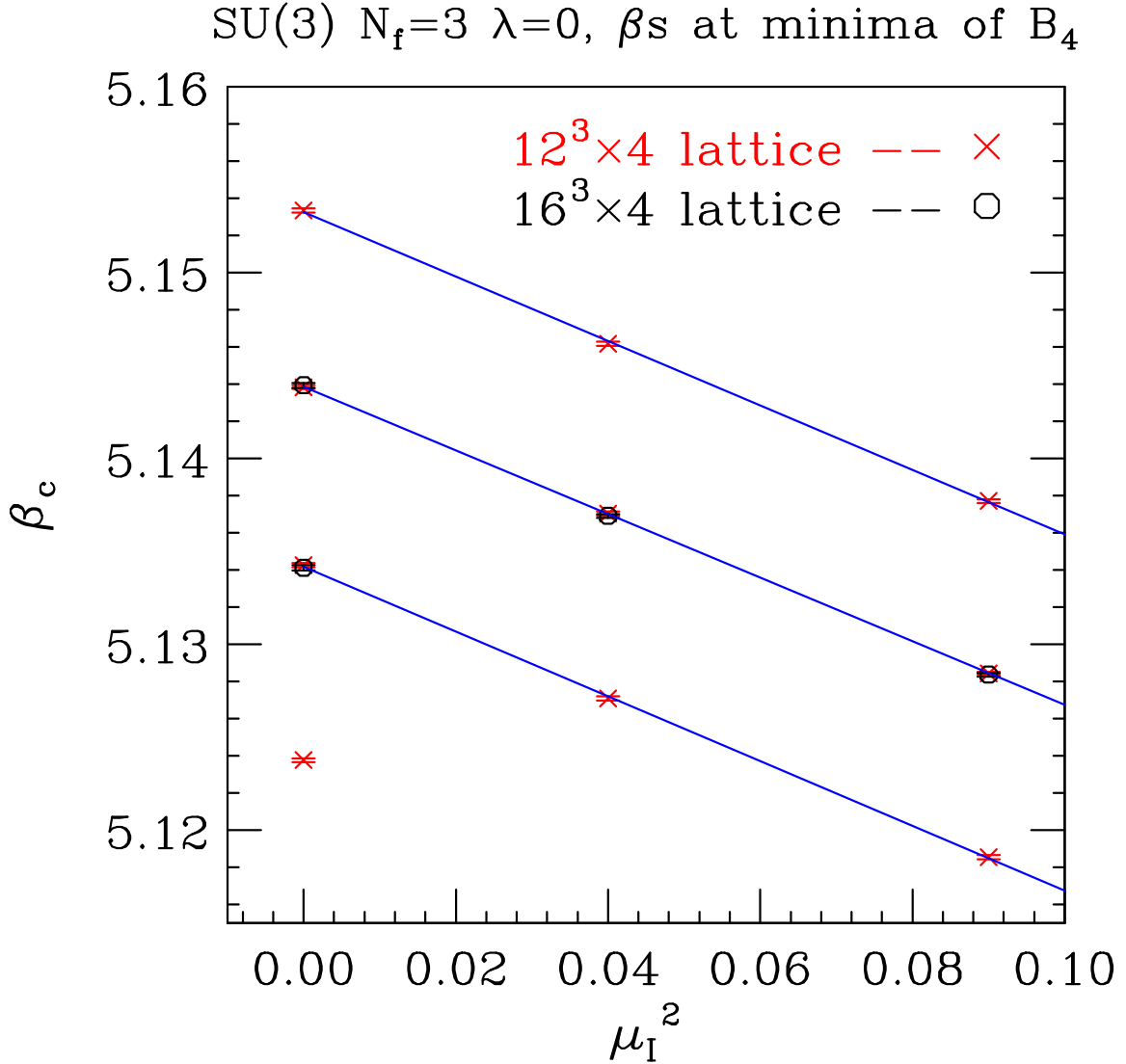}}
\caption{Transition $\beta$, $\beta_c$ as functions of $\mu_I^2$ for chosen
masses. The lines from top to bottom are for $m=0.035$, $m=0.03$ and $m=0.025$.
The isolated point is for $m=0.02$.}
\label{fig:beta_c}
\end{figure}

In figure~\ref{fig:beta_c} we plot the measured values of $\beta_c$ against
$\mu_I^2$ for each of the quark masses. Straight line fits appear adequate
with our current statistics. Although better fits could be obtained with a
$\mu_I^4$ term for $m=0.025$ and $m=0.035$ -- the $m=0.03$ straight line fit
is excellent -- the coefficients are clearly very small, and with only 3 points
on each curve, such an exact fit is hard to justify. These fits are to the
more extensive $12^3 \times 4$ `data'. We have plotted the $16^3 \times 4$
points on the same graph. These indicate that the finite size effects on 
$\beta_c$ are very small. For comparison with the work of others, these fits
are:
\begin{eqnarray}
\beta_c &=& 5.13418(10)   - 0.1743(18) \mu_I^2 \;\;\;\;\; m=0.025 \\
\beta_c &=& 5.14385(8)\;\:- 0.1711(13) \mu_I^2 \;\;\;\;\; m=0.030 \\
\beta_c &=& 5.15326(10)   - 0.1735(16) \mu_I^2 \;\;\;\;\; m=0.035 
\end{eqnarray}  
and $\beta_c=5.12377(10)$ at $m=0.02$, $\mu_I=0$.

\section{Reweighting studies}

As we saw in the previous section, the weak dependence of the Binder cumulants
on $\mu_I$, and the sizable statistical errors in determining this fluctuation
quantity mean that the observation that $B_4$ increases with $\mu_I$, while
strongly suggested is not definitive. Similar difficulties arise for simulations
at imaginary $\mu$. Here, de Forcrand, Kim and Philipsen have circumvented this
difficulty by calculating $\partial B_4 /\partial \mu^2$ directly 
\cite{deForcrand:2007rq}. They do this by calculating $B_4(\mu)$ and
$B_4(\mu+\delta\mu)$ in the same simulation. This is achieved by including the
ratio of determinants
\begin{equation}
\rho=\det[{\cal M(\mu+\delta\mu)}^{N_f/8}] /\det[{\cal M(\mu)}^{N_f/8}] 
\end{equation}
as a weight in the measurement of $\bar{\psi}\psi(\mu+\delta\mu)$ from the same
ensemble at chemical potential $\mu$ as is used to measure 
$\bar{\psi}\psi(\mu)$, namely
\begin{equation}
\langle\bar{\psi}\psi(\mu+\delta\mu)\rangle_{\mu+\delta\mu}
={\langle\rho\bar{\psi}\psi(\mu+\delta\mu)\rangle_{\mu} \over 
                                          \langle\rho\rangle_\mu},
\end{equation}
and similar expressions for higher powers of $\bar{\psi}\psi(\mu+\delta\mu)$.

Since exact calculation of such determinants is expensive, de Forcrand 
{\it et al.} used unbiased stochastic estimators for the ratio of determinants,
in particular,
\begin{equation}
\rho=\langle\exp[-\eta^\dag{\cal M(\mu)}^{-N_f/16}
{\cal M(\mu+\delta\mu)}^{N_f/16}{\cal M(\mu+\delta\mu)}^{N_f/16}
{\cal M(\mu)}^{-N_f/16}\eta + \eta^\dag\eta]\rangle_\eta
\end{equation}
where $\eta$ is Gaussian noise. The advantage of this method is that any
{\it finite} number of noise vectors gives an unbiased estimator of the
determinant. These authors reweighted from $\mu=0$ and performed a multistep
reweighting to $\mu=0.1$

\begin{figure}[htb]
\vspace{-0.25in}
\epsfxsize=5.5in
\centerline{\epsffile{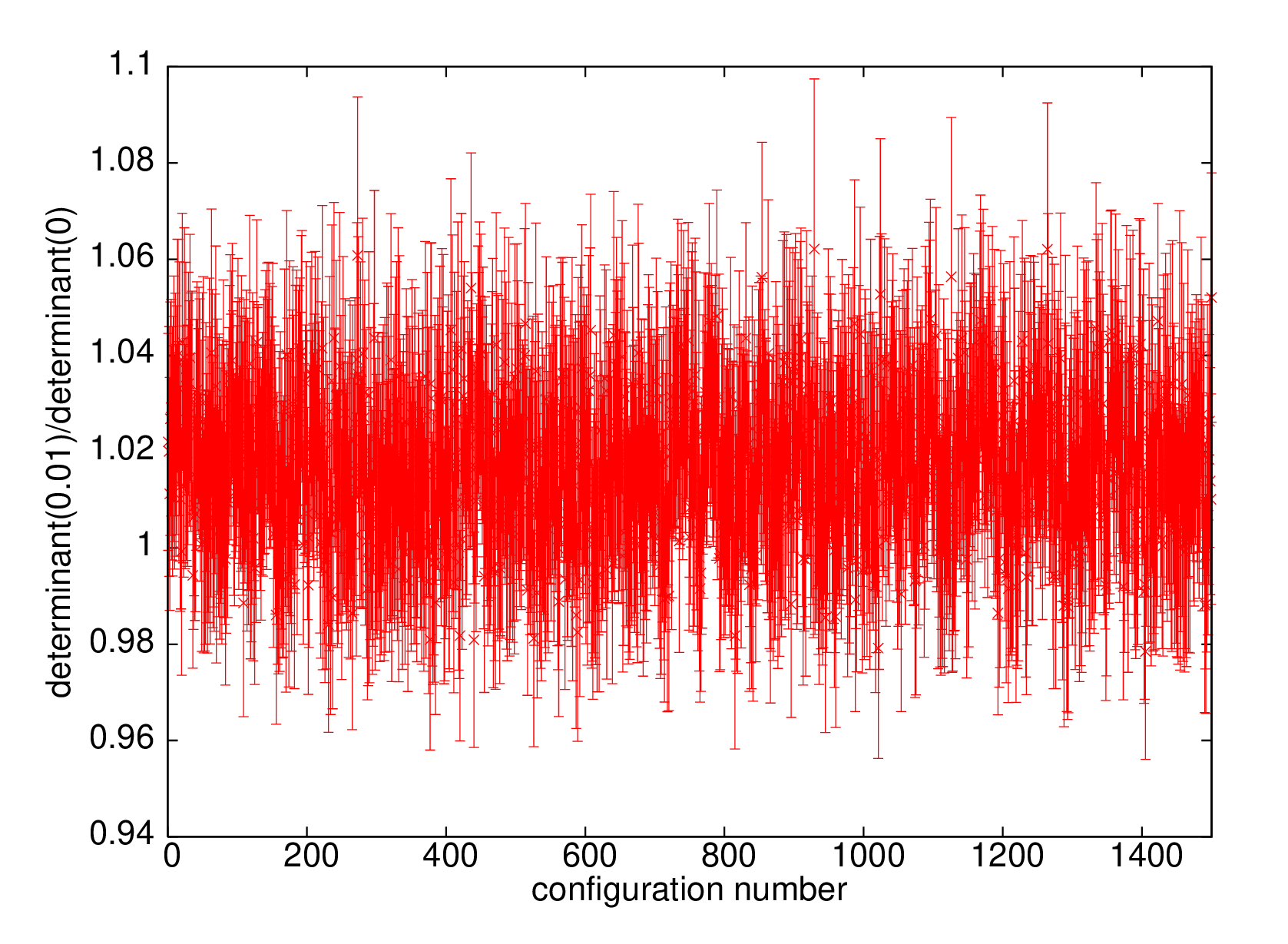}}
\vspace{0.25in}
\centerline{\epsffile{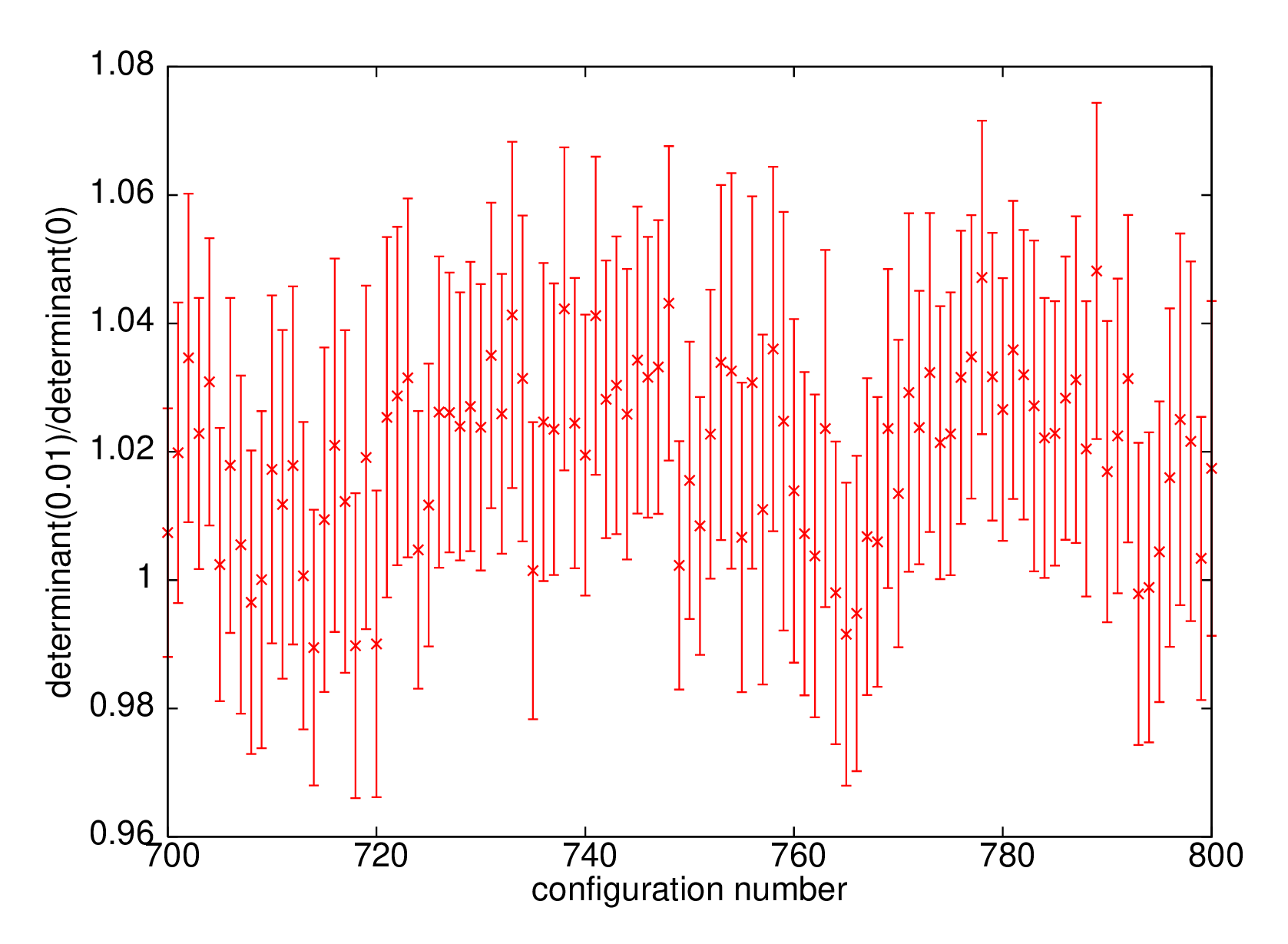}}
\caption{a) Stochastic estimates of the ratio of fermion determinants at 
$\mu_I=0.01$ and $\mu_I=0$ on a $12^3 \times 4$ lattice at $m=0.03$, $\mu_I=0$,
$\beta=5.143$. b) Section of graph (a) showing detail.}
\label{fig:det0.01_mu0}
\end{figure}

Whereas it appears that de Forcrand {\it et al.} limited themselves to 
$8^3 \times 4$ lattices, we are investigating applying this to $12^3 \times 4$
lattices, since the results of the previous section make it unclear whether
the slope $\partial B_4 /\partial \mu_I^2$ is the same for $8^3 \times 4$ 
lattices as it is for larger lattices. We first investigated the possibility
of reweighting from $\mu_I=0$ to $\mu_I=0.1$ in a single reweighting, but 
analysis of a few configurations quickly convinced us that although the
overlap might be reasonable, the fluctuations were so large as to make it
impossible to obtain a reasonable estimate of the determinant without use of
far more noise vectors than is reasonable. We then went back to a
reweighting from $\mu_I=0$ to $\mu_I=0.01$ as a single step process, and one
that could be used as a basis for a multistep reweighting to an even larger
$\mu_I$. For this trial run we used 1500 configurations at $m=0.03$, separated 
by 200 trajectories. For each configuration we used 200 noise vectors with 
$\delta\mu_I=0.01$ and the same set of noise vectors with $\delta\mu_I=-0.01$,
making use of the fact that the determinant for a single configuration 
remains unchanged under $\mu_I \rightarrow -\mu_I$ to remove 
${\cal O}(\delta\mu_I)$ fluctuations in our noisy estimator. We used 1000 noise
vectors for our noisy estimators for $\bar{\psi}\psi(0)$ and the same set
for $\bar{\psi}\psi(0.01)$ and $\bar{\psi}\psi(-0.01)$. This effectively
removes the errors in using noisy estimators of the condensate from 
consideration. The resulting estimate of $\partial B_4 /\partial \mu_I^2$ is
$3.1 \pm 4.1$ compared with the estimate $0.38 \pm 0.22$ obtained in the
previous section. This indicates that, as expected, we need to use a 
$\delta(\mu_I^2)$ much greater than the $0.0001$ used here, which will 
require a multistep reweighting in order to avoid large fluctuations. 
Our estimate for $\partial\beta_c/\partial\mu_I^2$ is $-0.177(9)$, in agreement
with $-0.171(1)$ obtained in the previous section. Figure~\ref{fig:det0.01_mu0}
shows the noisy estimators of the determinants with errors that we obtained.
We see that the errors are comparable with the difference of these determinant
ratios from one and from their mean, which is one reason why the signal/noise 
ratio is so poor. 

\begin{figure}[htb]
\vspace{-0.25in}
\epsfxsize=5.5in
\centerline{\epsffile{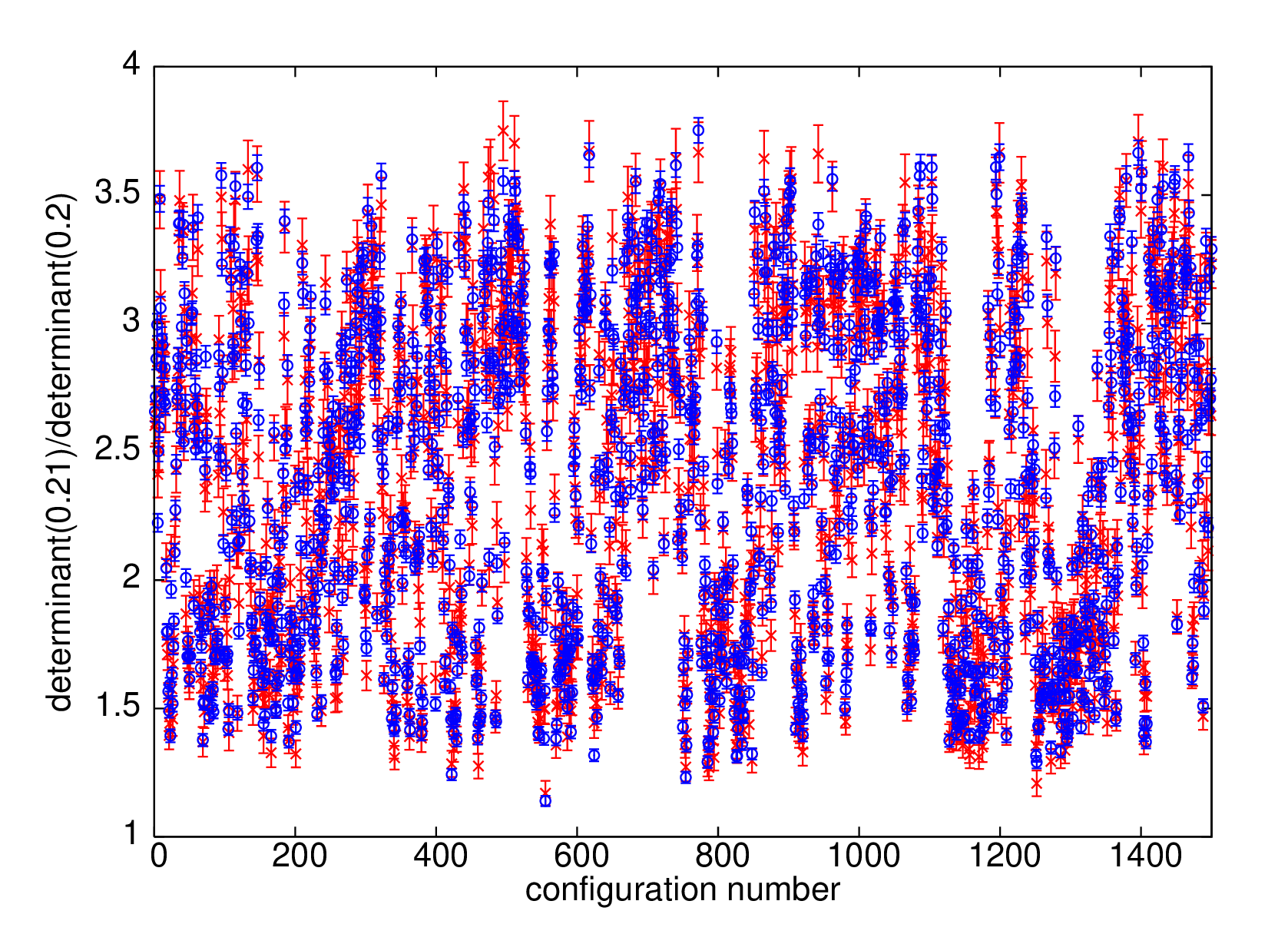}}
\vspace{0.25in}
\centerline{\epsffile{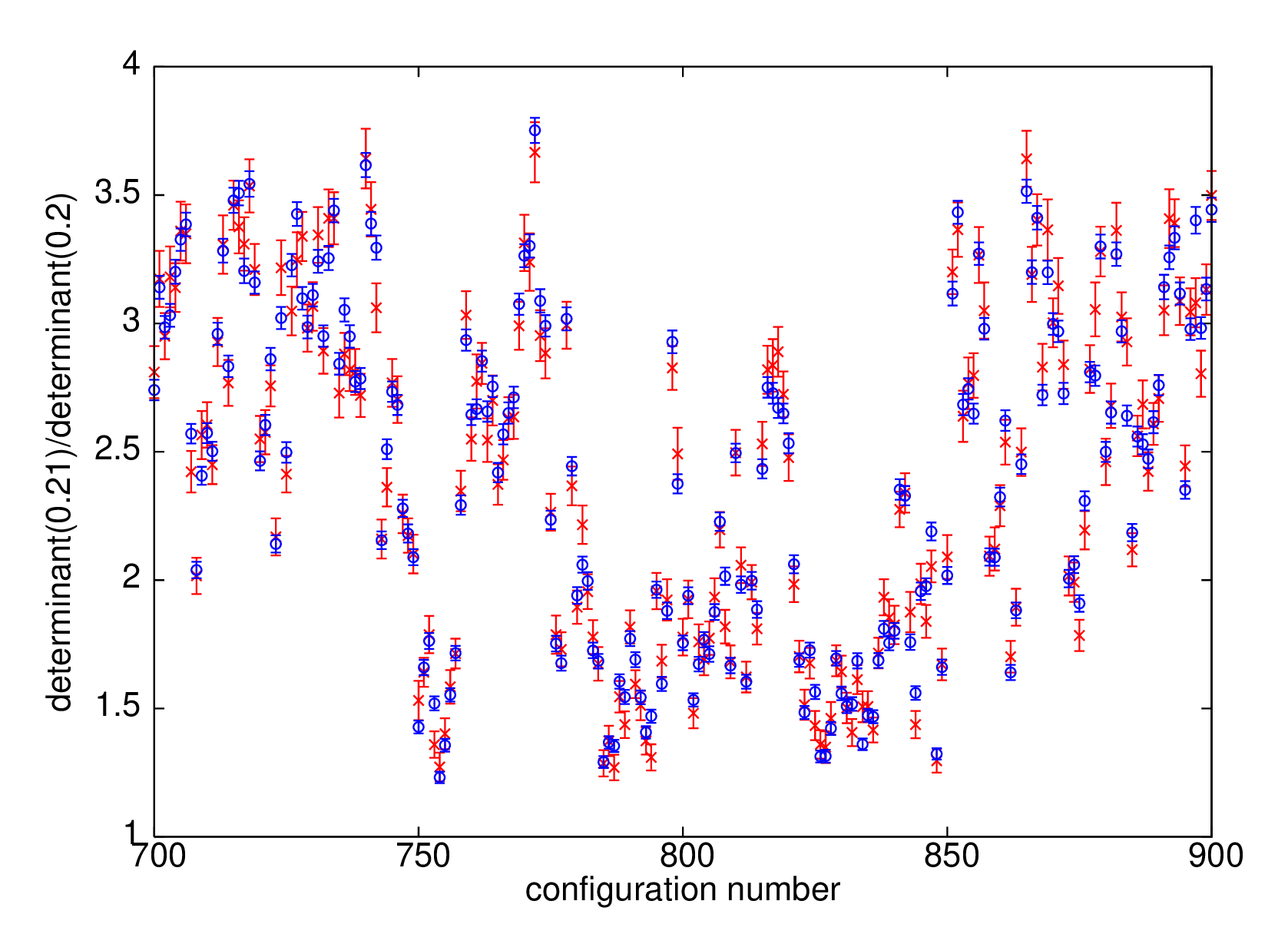}}
\caption{a) Stochastic estimates of the ratio of fermion determinants at
$\mu_I=0.21$ and $\mu_I=0.2$ on a $12^3 \times 4$ lattice at $m=0.03$, 
$\mu_I=0.2$, $\beta=5.137$. The crosses(red online) are for 200 noise vectors;
The circles(blue online) are for 1000 noise vectors. 
b) Section of graph (a) showing detail.}
\label{fig:det0.01_mu2}
\end{figure}

We notice with the reweighting from $\mu_I=0$, that one problem is that the
signal is of order $\delta(\mu_I^2)=(\delta\mu_I)^2$, while the noise is of
order $\delta\mu_I$. While this can be overcome with a multistep (multiple
$\mu_I$s) reweighting, an alternative way of avoiding this difficulty is to 
start at non-zero $\mu_I$ where for small $\delta\mu_I$, $\delta\mu_I$ and 
$\delta(\mu_I^2)$ are of the same order of magnitude. We have thus tried 1-step
reweighting from 1500 configurations at $\mu_I=0.2$, with $\delta\mu_I=0.01$
and hence with $\delta(\mu_I^2)=0.0041$. For this test we ran first with 200
noise vectors for each configuration, and later with 1000 noise vectors for
each configuration. 1000 noise vectors were used in estimating the chiral
condensate. Using 200 noise vectors to estimate the determinant ratio we
obtained $\partial B_4 /\partial \mu_I^2 = -0.39 \pm 0.56$ and for 1000 noise 
vectors $-0.54 \pm 0.45$. Although this indicates that we still do not have
enough statistics, we would only need to reduce the statistical errors by an
order of magnitude to make a definitive prediction. 
$\partial\beta_c/\partial\mu_I^2$ measured in the same calculations is
$-0.175(2)$ compared with $-0.171(1)$ calculated in the previous section. In
figure~\ref{fig:det0.01_mu2}, we show our estimates of the ratio of fermion
determinants.
Even with 200 noise vectors/configuration, the ratio of determinants is well
determined. The statistical errors are considerably smaller than the ratio's
departure from unity and, more importantly, considerably smaller than the range
of values taken by this ratio over the ensemble of configurations. This
presumably is why little improvement in the estimate of $\partial B_4/\partial
\mu_I^2$ is obtained by increasing the number of noise vectors from 200 to
1000. Comparison of the estimates of the determinant ratios for 200 and 1000
noise vectors makes us confident that our noisy estimates are reliable.

In case the main problem was overlap, we reduced our $\delta\mu_I$ to $0.005$.
The results, however, were similar. $\partial\beta_c/\partial\mu_I^2$ was well
determined, while the errors in $\partial B_4/\partial \mu_I^2$ exceeded the
signal. The determinant ratio was well determined by 200 noise vectors, the
errors being much smaller than the fluctuations in the value of this ratio
from configuration to configuration.

To get an accurate estimate of $\partial B_4 /\partial \mu_I^2$ will require
the analysis of many more configurations. These configurations should be 
separated by enough trajectories as to make them reasonably independent or
they are unlikely to improve the errors, just as increasing the number of
noise vectors for estimating each ratio from 200 to 1000 did not significantly
improve our errors. In fact, using 1 noise-vector to estimate the determinant
ratio at the end of each of 300,000 consecutive trajectories at $\mu_I=0.2$,
gave similar accuracy to using 200 or 1000 noise-vectors for each of our 1500
configurations spaced by 200 trajectories. In addition, producing a single
trajectory takes much less computer time than a measurement with (say) a 100
noise-vector estimate of the determinant ratio and a 100 noise-vector estimate
of $\bar{\psi}\psi$. In addition we need to check whether a single- or a
multi-step estimation of the determinant ratio is more efficient, even at
$\mu_I=0.2$ where it is not forced on us by other considerations.

\section{Discussions and Conclusions} 

We have studied the finite temperature transition for 3-flavour lattice QCD
with a finite chemical potential $\mu$ in the phase-quenched approximation,
where the phase of the fermion determinant is set to zero, using RHMC 
simulations. This can be considered as studying lattice QCD with $3/2$ up-type
quarks and $3/2$ down-type quarks at a chemical potential $\mu_I=2\mu$ for
isospin ($I_3$). As we have indicated in the introduction, in the small
$\mu$($\mu_I$) regime -- $\mu_I < m_\pi$ -- there are indications that the
dependence of the transition temperature on $\mu$ for the phase-quenched model
is similar if not identical to that for full QCD. Within the limitations of
our statistics, which only allow us to include terms linear in $\mu^2$, we find
that the coefficient of $\mu^2$ in the fit to $\beta_c(\mu^2)$, is within 10\%
and probably within 5\% of that obtained by de Forcrand and Philipsen for the
full theory by continuation from imaginary $\mu$ 
\cite{deForcrand:2006pv,deForcrand:2007rq}. This gives further evidence
that the $\mu^2$ dependence of $T_c$ is the same in phase-quenched and full
QCD. If this is true, it is reasonable to expect that the nature of the 
transition will be the same in both theories. 


Let us briefly digress to discuss other recent work which could have relevance
to the connection between phase-quenched and full QCD. Some recent work of
Fodor, Katz and Schmidt, which employs the density-of-states method, uses the
phase-quenched theory as a starting point for their factorized reweighting
\cite{Fodor:2007vv}. This shows a small but finite shift in $\beta_c$ in
reweighting from phase-quenched to full QCD. However, all the $\mu$ values which
they consider are larger than $m_\pi/2$, and so in the region where the two
theories are no longer expected to be similar. There has also been extensive
work on QCD at finite chemical potentials using a random matrix/chiral
perturbation theory approach \cite{Splittorff:2007ck,Splittorff:2007zh}. This
has indicated that the phase of the fermion determinant becomes much worse
behaved at for $\mu > m_\pi/2$. Although this work does not (yet) explain why
the full and phase-quenched QCD  behave similarly, it does indicate that the 
same pion modes describe the physics of each, and suggests model calculations 
which might clarify the situation.

It was expected that the critical point at zero chemical potential would
move to higher mass at finite chemical potential. If so, for quark masses
just above the critical mass at $\mu=\mu_I=0$, this would become the 
sought-after critical endpoint where the crossover at $\mu=0$ would change to
a first-order transition. Our simulations for $m$ close to $m_c(0)$ indicate
that this does not happen, but rather $m_c(\mu_I)$ decreases with increasing
$\mu_I$. The $\mu_I$ dependence of the Binder cumulant used to determine the
nature of the transition is very weak for the lattice sizes we use 
($8^3 \times 4$, $12^3 \times 4$ and $16^3 \times 4$). For this reason, our
results can only be considered suggestive, and not definitive. Similar
conclusions have been drawn by de Forcrand {\it et al.} from simulations at
imaginary $\mu$ \cite{deForcrand:2006pv,deForcrand:2007rq}. This disagrees
with the early work of the Bielefeld-Swansea collaboration
\cite{Karsch:2003va}, who did claim to find such a critical endpoint. However,
as indicated before, these simulations used the R algorithm which as 
de Forcrand and Philipsen and we discovered can lead to misleading results.

De Forcrand {\it et al.} have recently introduced reweighting methods which
enabled them to calculate the slope of the Binder cumulant directly, thus
reducing the errors to a point where the sign is determined unambiguously
\cite{deForcrand:2007rq}. This shows that the critical mass does indeed
decrease with increasing $\mu$, so that there is no critical endpoint
associated with $m_c(0)$. However, their published results using this new
method are all on $8^3 \times 4$ lattices where finite size effects, such as
the fact that the chiral condensate is not the true order parameter (in the
renormalization group sense), are large.

For this reason we have been investigating the use of such reweighting 
techniques for phase-quenched QCD on $12^3 \times 4$ lattices. Larger lattices
are less suited to such reweighting because the overlap between the ensembles
of configurations at $\mu_I$ and $\mu_I+\delta\mu_I$ for given $\mu_I$ and
$\delta\mu_I$ is smaller for larger lattices. The ratio of determinants is
further from unity for the larger lattices, and the fluctuations associated
with the noisy estimator on a single configuration are also larger. 
Correlations in molecular-dynamics time are longer on the larger lattice. Our
tests are promising and suggest that using finite rather than zero $\mu_I$
configurations for the reweighting are preferable. However, since reweighting
is expensive, unless we can find a way to make better use of the fact that
the ratio of fractional powers of Dirac operators for $\mu_I$ and 
$\mu_I+\delta\mu_I$ is better conditioned than either of the original 
operators, reweighting will be considerably more expensive than the cost of
producing a single trajectory, so that it is unclear as yet whether 
reweighting will prove to be the most cost-effective method of getting
definitive results on these larger lattices. 

One might ask whether our failure to find a critical endpoint disagrees with
the work of Fodor and Katz \cite{Aoki:2006we}. They reported a critical
endpoint at $\mu_B=360(40)$~MeV and $T=162(2)$~MeV, and hence
$\mu_I=240(27)$~MeV. Since our method breaks down for $\mu_I \gtrsim m_\pi
\approx 140$~MeV, their value is beyond the reach of our method. Hence our
simulations do not show the absence of a critical endpoint, only the absence
of a critical endpoint associated with the critical point at $\mu=\mu_I=0$,
for 3-flavour QCD.

We are now extending our simulations of 3-flavour phase-quenched lattice QCD to
enable a calculation of the equation-of-state of this theory outside the
superfluid region. This will enable comparison with full QCD. In addition,
the phase-diagram of QCD at finite isospin chemical potential and its
equation-of-state are of interest in their own right. This has led to new
activity in the studies of these theories \cite{deForcrand:2007uz}.

All our simulations have been performed with the standard staggered action,
with $N_t=4$ and are thus subject to large discretization errors. The
Bielefeld-Swansea collaboration found that the critical mass at $\mu=0$
decreased dramatically, when they changed their lattice action from the
standard lattice action to a highly improved action, indicating that this mass
is very sensitive to finite lattice-spacing errors
\cite{Karsch:2003va,Schmidt:2004ke}. A less dramatic decrease in the critical
mass has recently been reported by de Forcrand, Kim and Philipsen when they
increased $N_t$ from $4$ to $6$ with the standard staggered action
\cite{deForcrand:2007rq}. Hence we should consider repeating our simulations
at larger $N_t$, improving the action we use, or both.

In using a staggered action for 3 flavours, we are ignoring the so-called
`rooting' controversy. People have questioned whether taking fractional powers
of the fermion determinant to allow use of staggered fermions to simulate
numbers of fermion flavours which are not multiples of 4, defines a theory
with a sensible continuum limit. We direct the reader to
\cite{Creutz:2007yr,Bernard:2007ma} for recent arguments on both sides of this
controversy. Even assuming that this controversy is resolved and indicates
that `rooted' staggered fermions are legitimate, this is only relevant to the
case of zero $\mu$. At $\mu\ne 0$, Golterman, Shamir and Svetitsky have
pointed out that further ambiguities arise with regard to taking fractional
powers of the phase of the determinant \cite{Golterman:2006rw}. We have
avoided this difficulty by ignoring the phase, but eventually it will need to
be faced.


\section*{Acknowledgements}

We thank Ph.~de~Forcrand for his help and for useful discussions. We also thank
O.~Philipsen and F.~Karsch for helpful discussions. The simulations reported
here were performed on Jacquard and Bassi at NERSC on an ERCAP allocation and
on Tungsten, Copper, Abe and Cobalt at NSCA and DataStar at SDSC under an NRAC
grant.

\end{document}